\documentclass[fleqn,usenatbib,useAMS]{mnras}

\usepackage{graphicx}	
\usepackage{amsmath}	
\usepackage{amssymb}	
\usepackage{multicol}        
\usepackage{bm}		
\usepackage{pdflscape}	

\usepackage{tablefootnote}
\usepackage{makecell}   
\usepackage{color} 
\usepackage[utf8]{inputenc}
\usepackage{float}

\usepackage[T1]{fontenc}
\usepackage{ae,aecompl}
\usepackage{newtxtext,newtxmath}
\usepackage{subcaption}
\usepackage{multirow}
\usepackage{caption}

\title[The age sequence of young clusters in Perseus]{The age sequence of young clusters in Perseus: Estimating ages from mass distributions}

\author[T. Pavlidou et al.]{
Tatiana Pavlidou,$^{1}$ 
Aleks Scholz,$^{2}$\thanks{E-mail: as110@st-andrews.ac.uk}
Koraljka Muzic $^{3}$\\
$^{1}$ Department of Engineering, Cyprus University of Technology, 30 Arch. Kyprianos Str.
3036, Limassol, Cyprus \\
$^{2}$ SUPA, School of Physics \& Astronomy, University of St Andrews, North Haugh, St Andrews KY16\,9SS, United Kingdom\\
$^{3}$ Instituto de Astrofísica e Ciências do Espaço, Faculdade de Ciências, Universidade de Lisboa, Ed. C8, Campo Grande, 1749-016
Lisbon, Portugal \\
}

\date{submitted: 01 Sep 2025; accepted: 20 Jan 2026}

\pubyear{2025}

\begin{document}
\label{firstpage}
\pagerange{\pageref{firstpage}--\pageref{lastpage}}
\maketitle

\begin{abstract}
Establishing ages for young clusters is key for properly tracking the star formation history of a region. In this paper we investigate a new approach to estimating ages for young populations, based on the well-founded assumption that the initial mass function is the same throughout a star forming cloud. We trial this method for six young clusters in the Perseus star forming region. For all six clusters, we construct new member samples in a homogeneous way using \textit{Gaia} DR3. We estimate masses by comparing \textit{2MASS} photometry to theoretical isochrones, including Monte Carlo simulations to propagate the errors. We compare the mass distributions of the clusters for a range of plausible ages, looking for a combination of ages that results in indistinguishable mass distributions across the region. We find the best fit for ages of 1\,Myr for NGC\,1333+Autochthe, 2\,Myr for IC\,348, 2-3\,Myr for Heleus, 3-4\,Myr for Mestor, 4-5\,Myr for Electryon+Cynurus, and 5-8\,Myr for Alcaeus. All other combinations of ages are ruled out by this criterion. The established age sequence is consistent with the relative ages inferred from disc fractions, and broadly aligns with the age sequence determined in previous studies using isochrone fitting. We suggest that this approach can be a useful complement and cross-check to established methods to estimate ages in young populations.
\end{abstract}

\begin{keywords}
stars: formation -- stars: general --, stars: kinematics and dynamics -- stars: luminosity function, mass function -- stars: pre-main-sequence
\end{keywords}



\begingroup
\let\clearpage\relax
\endgroup
\newpage

\section{Introduction}
\label{sec:intro}

\begin{figure*}[t]
    \includegraphics[width=\textwidth]{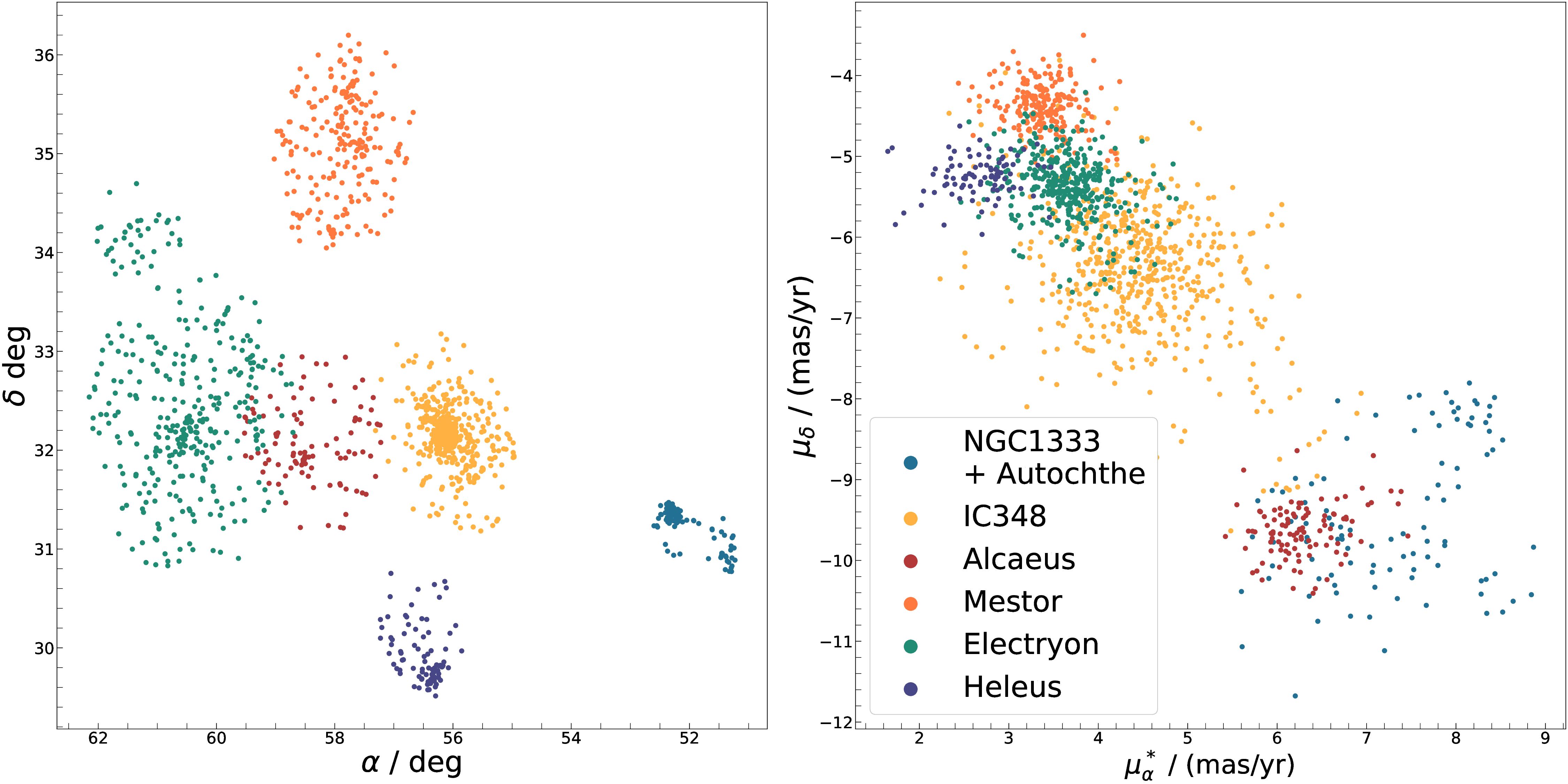}
    \caption{The final members of the clusters in spatial distribution (left) and in proper motion (right).}
    \label{fig:clusters}
\end{figure*}

Star clusters provide us with a snapshot of the properties of stars at a specific point in their evolution. To establish an evolutionary sequence, it is critical to determine ages of clusters, at the minimum in a relative sense. For the main sequence, there are a number of sophisticated methods for dating clusters, including gyrochronology, isochrone fitting, measuring the Lithium depletion boundary, or activity measurements. For ages younger than 10\,Myr, all these methods fail or become unreliable to some degree. Measuring robust stellar ages for clusters in the earliest evolutionary stages is a key problem in star formation (see review by \cite{soderblom_2014}).

As young stars evolve towards the main sequence, they contract and thus move downwards in the HR diagram. Therefore, the mass-luminosity relation changes significantly as a function of age. As a result, adopting a plausible age is critical when determining the mass function of young clusters. In this paper we propose to turn this problem around: Instead of adopting an age to determine the mass distribution we suggest to determine ages assuming a universal mass function.

The universality of the local mass function has now been established without reasonable doubt from a large number of young clusters in the galactic neighborhood (see review by \cite{bastian_2010}). While occasional deviations from universality have been pointed out, these have typically proven spurious with further analysis and/or more complete surveys. Most recently, \cite{damian_2021} have determined and compared mass functions for 8 clusters spanning a range of properties (distances, UV radiation, stellar density). They do not find strong evidence for environmental differences on the mass function. Star forming young clusters are also unaffected yet by evolutionary effects, such as mass segregation. While the mass function in the substellar domain is still under debate, the clusters surveyed with sufficient depth also seem to harbour similar numbers of brown dwarfs (relative to those of stars, see \cite{muzic_2017} for an overview). In this paper we assume that the mass function is the same throughout a specific star forming region to estimate the ages of the clusters in that region.

The star forming complex in Perseus is long known to harbour two active star forming clusters, namely NGC\,1333 and IC\,348 (see \cite{bally_2008} for a review). In addition, \cite{pavlidou_2021} find 5 new substantial clusters, named Autochthe, Alcaeus, Mestor, Electryon, and Heleus, grouped around the two core clusters. Two further studies identify a couple of other groups named Cynurus \citep{kounkel_2022} and Gorgophone \citep{olivares_2023}. The latter paper also determines ages and masses for young stars in this region, concluding that the mass distributions are compatible with the Chabrier mass function. In this paper we build on these previous studies and provide an independent estimate of the age sequence -- and thus star forming history -- of the Perseus clusters. In Section \ref{sec:samples} we determine new membership lists for these clusters in a homogeneous way. In Section \ref{sec:massdist} we determine the mass distributions for our samples. We use these distributions to assess the ages of the clusters, in Section \ref{sec:ages}. Finally, in Section \ref{sec:discussion} we summarise and contextualise our findings. 

\section{Selection of cluster samples}
\label{sec:samples}

\begin{figure*}
    \includegraphics[width=\textwidth]{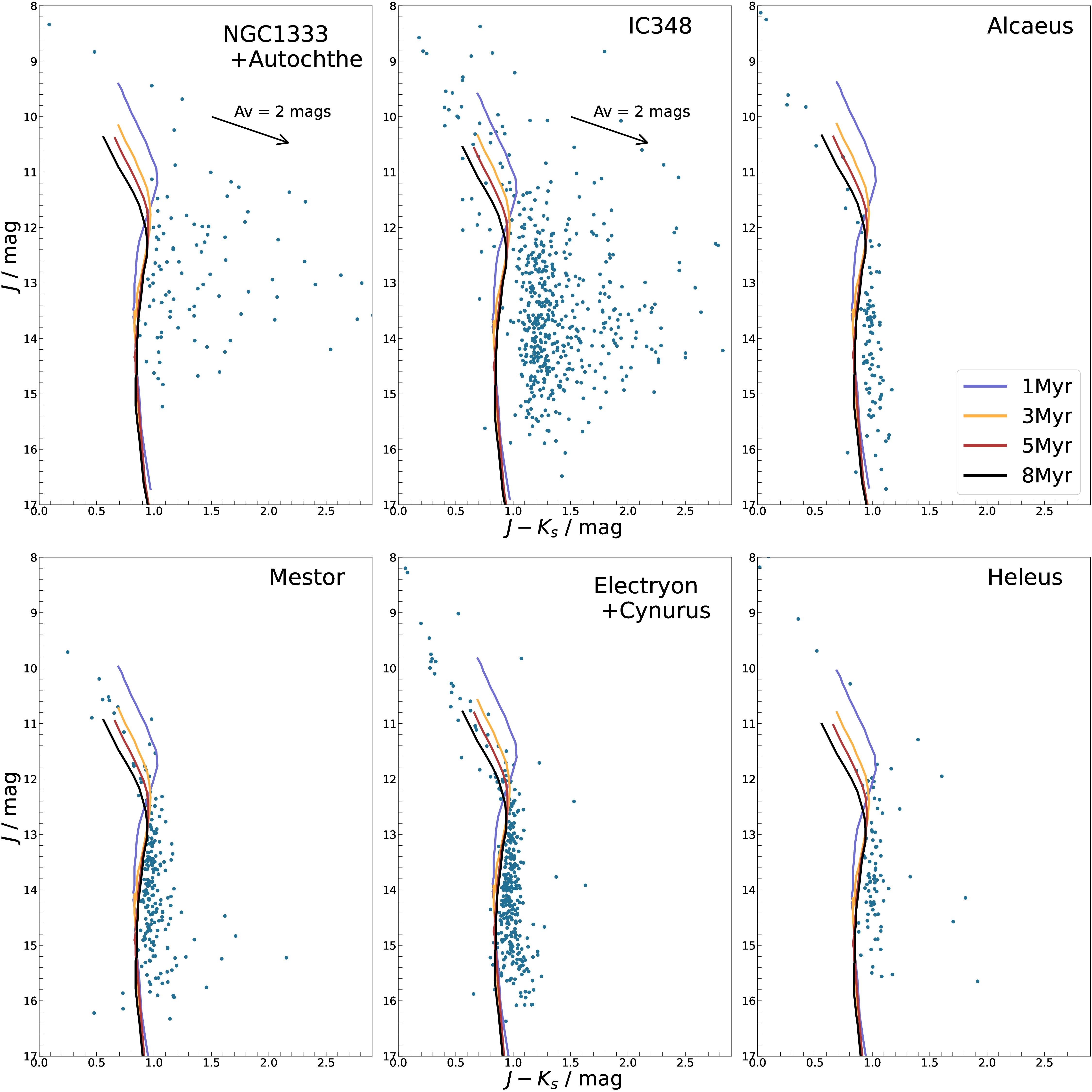}
    \caption{CMDs from \textit{2MASS} photometry for all clusters. Isochrones shown (same in all panels) are from \citet{baraffe_2015} and are corrected for the distance to each cluster. The arrow depicts an extinction equivalent to 2.0 mags. For clarity the range in both axes is the same in all panels.}
    \label{fig:cmds}
\end{figure*}

\begin{figure}
    \includegraphics[width=\columnwidth]{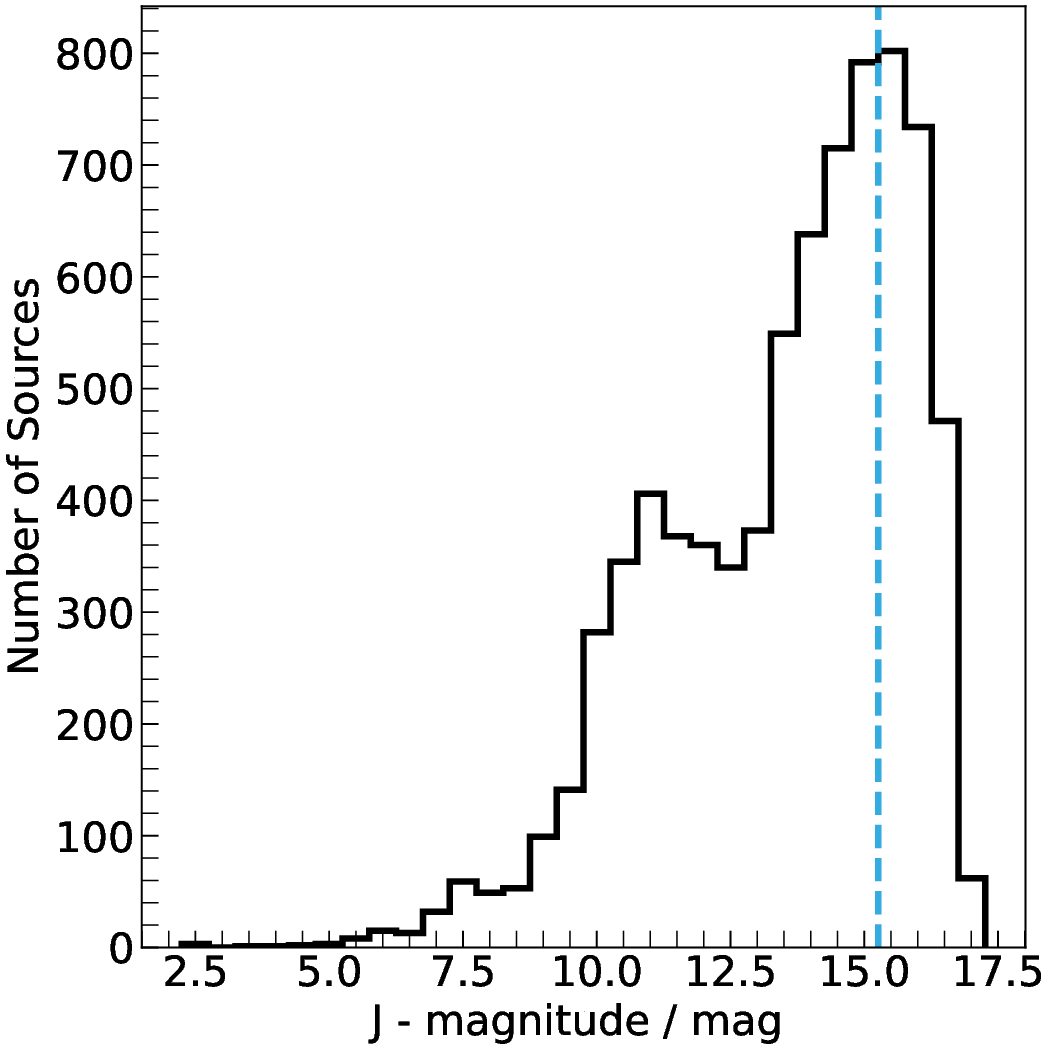}
    \caption{Histogram of the J-magnitudes of the whole initial Perseus sample (7716 matches in \textit{2MASS} out of 8123). The cyan dashed line marks the completeness limit of the sample.}
    \label{fig:j_magnitudes}
\end{figure}

\begin{table}
\centering
\caption{Initial Constraints for the Perseus Sample}
\begin{tabular}{l c} 
\hline
\hline
Property & Condition \\ [0.5ex] 
\hline
Sky Coverage (deg) & $45 < \alpha < 64$, $24 < \delta < 40 $ \\
$\mu_{\alpha}^*$ (mas\,yr$^{-1}$) & $-5.0<\mu_{\alpha}<13.0$ \\ 
$\mu_{\delta}$ (mas\,yr$^{-1}$)  & $-15.0<\mu_{\delta}<-0.5$ \\
Parallax Angle (mas) & $  1.8<\varpi<4.3$ \\
Parallax Angle error (mas) & $\sigma_{\pi} < 0.7$ \\
Approximate Age (Myr) & $\lesssim 10 $ \\ 
Number of Sources &  8123 \\
\hline
\hline
\end{tabular}
\label{tab:perseus_conditions}
\end{table}

We define an initial \textit{Gaia} DR3 \citep{Gaia_2016,Gaia_2023} sample for the Perseus region satisfying the conditions shown in Table \ref{tab:perseus_conditions}. The sky coverage is selected such that it includes all known clusters in the region (see Section \ref{sec:intro}). The proper motion, parallax and parallax error conditions adopted here are motivated by the range covered by the confirmed members in IC\,348 and NGC\,1333 as listed in the comprehensive photometric and spectroscopic survey by \cite{luhman_2016}. Finally, we apply an age cut, keeping sources younger than $\sim$10\,Myr by comparing with an isochrone by \cite{marigo_2017}. The resulting sample for the Perseus region consists of 8123 sources. This represents a preliminary sample that is intended to screen out the majority of the background.

To build samples for the six Perseus clusters in a consistent manner we use parallax, proper motion, and spatial positions as primary criteria to select cluster members. In the following we treat NGC\,1333 and Autochthe as one cluster due to the similarities they exhibit across the parameter spaces of proper motion, parallax and age \citep{pavlidou_2021}. We also combine Cynurus together with Electryon, similarly motivated by the fact that they overlap in proper motion and parallax \citep{kounkel_2022}. 

To select members for each cluster we adopt a similar procedure as followed in \cite{pavlidou_2021}, but now using data from \textit{Gaia} DR3. We start with an initial sample for each cluster defined by a box in the spatial distribution. The box centers and half-lengths for each cluster are listed in the Appendix \ref{sec:selection_appendix} and are motivated by previous studies on the clusters \citep{pavlidou_2021, kounkel_2022, pavlidou_2022}. We then calculate the mean positions ($\alpha$,$\delta$)$_\mathrm{mean}$) and standard deviation ($\sigma(\alpha$,$\delta)$) of this sample using the \textit{fit\_bivariate\_normal} package from \textit{astroML.stats} and \textit{bivariate\_normal} from \textit{astroML.stats.random} in \textsc{PYTHON}.  We use a 3$\sigma$ ellipse in spatial distribution to select members for NGC\,1333/Autochthe (treated as one cluster), and for IC\,348. For the remaining clusters, a 2$\sigma$ ellipse turned out to be more useful.

We carry out a similar selection procedure in proper motion space for each cluster, keeping members within the 3$\sigma$ ellipse for NGC\,1333/Autochthe, 5$\sigma$ for IC\,348, Alcaeus and Cynurus and 6$\sigma$ for Mestor, Electryon and Heleus. We iteratively vary the radius of the ellipse and check the spatial, proper motion, parallax and colour-magnitude distributions (CMDs) of the sample such that the sources are clustered around a prominent mean value in all of these parameters via visual inspection. Our final choice of the selection criteria is a result of this iterative process. We cross-match our \textit{Gaia} samples with the 2MASS Survey \citep{cohen_2003} using a 1.0'' tolerance window. Almost all selected stars have a 2MASS entry.

We include the detailed information on the cluster parameters in Appendix \ref{sec:selection_appendix}. Figure \ref{fig:clusters} shows our final samples for all clusters in spatial and proper motion distributions. Figures \ref{fig:cmds} shows CMDs using 2MASS photometry along with isochrones from \cite{baraffe_2015}. The typical uncertainties in $J$ and $K$ magnitudes are $\sim 0.03$\,mags. We note that these samples are very similar to the ones defined in \citet{pavlidou_2021}. In Table \ref{tab:cluster_samples} we summaries the numbers in our newly selected samples.

To check for possible contamination in our samples, we repeated the process of the cluster selection, at a position in our field without a clearly visible cluster. When doing so, we do not see a clustering in proper motion or parallax, as expected, confirming that our samples are representative of actual clusters. For arbitrary positions in spatial distribution and proper motions as well as field sizes similar to the one chosen for our clusters, we typically end up with very few objects (less than 10) that also fall into a narrow parallax range. To re-iterate, we start our selection with an initial sample that already satisfies broad selection criteria based on proper motion, parallax, and age, and therefore should screen out most of the background and foreground. Altogether this simple test demonstrates that the contamination of our samples is considered to be low.

In the following sections we also test the effect that binary sources may have on our results. We therefore also construct samples that exclude potential binaries using the RUWE statistic provided in \textit{Gaia}. For each cluster we remove sources with RUWE $> 1.4$ \citep{lindegren_2018} which eliminates at most a couple of dozens of stars. In Table \ref{tab:cluster_samples} we show our initial \textit{Gaia} samples, our \textit{2MASS} samples and our \textit{2MASS} samples excluding binaries.

To determine the completeness of our samples we explore the distribution of the $J$ magnitudes of our initial Perseus sample. We adopt as the completeness limit the $J$ magnitude at the peak of the histogram which is $\sim15.3$\,mags (see Figure \ref{fig:j_magnitudes}). Converting this magnitude to the equivalent mass limit depends on the age, distance and extinction correction used. We adopt an average distance of 300\,pc and an average extinction in $A_V$ of 1\,mags. Because age affects the conversion of $J$ magnitude to mass significantly, we estimate the mass completeness limit according to each of the ages assumed in our following analysis (see Section \ref{sec:ages}).

\begin{table}
\centering
\caption{Cluster samples in this work. The number of objects identified from Gaia DR3, and the counterparts in 2MASS, and the number with low ruwe value (see text).} 
\label{tab:cluster_samples}
\begin{tabular}{l | c c c c}
\hline\hline
Cluster  & \textit{Gaia} DR3 & \textit{2MASS}  & with ruwe $<$ 1.4  \\
\hline
IC\,348                 & 549  & 539  & 505  \\
NGC\,1333+Autochthe     & 100  & 94   & 92   \\
Alcaeus                 & 103  & 101  & 92   \\
Mestor                  & 220  & 217  & 206  \\
Electryon+Cynurus       & 364  & 355  & 333  \\
Heleus                  & 89   & 88   & 85   \\
\hline\hline
\end{tabular}
\end{table}

\section{Masses and mass distributions}
\label{sec:massdist}

In this section we derive the distribution of masses for our samples of cluster members. To re-iterate our approach from Section \ref{sec:intro}, our goal here is to define the ages of the clusters, under the assumption that the mass distributions should be constant across the Perseus star forming complex.

When estimating masses, the distance and age enter as free parameters. The distances to the clusters studied here are well defined thanks to the available Gaia data. In the following, we use the average parallax for our selected samples, converted to distance. Those values are listed in Table \ref{tab:cluster_conditions} in the Appendix. We assume a range of plausible ages for each cluster, from 1 to 5\,Myr in steps of 1\,Myr, plus 8\,Myr, which are the available age steps in \cite{baraffe_2015}.

The general procedure we use to estimate masses follows closely the methodology outlined in \cite{muzic_2019}, which we summarise below. We first derive the extinction, by calculating the length of the reddening vector between the object's position in $(J,J-K)$ space (Figure \ref{fig:cmds}) and the isochrone for a given age and distance from \citet{baraffe_2015}. For the reddening path, we use the relations published in \cite{wang_2019}, which are displayed in the first two panels of their Figure \ref{fig:cmds}. We note that for most sources outside NGC\,1333 and IC\,348 the extinction is either zero or negligible. For a few objects on the blue side of the isochrone we assume an extinction of zero. We then compare the dereddened \textit{2MASS} photometry with the isochrone to find the best fitting mass. This includes an interpolation between the discrete mass intervals available in the isochrone.

In order to incorporate the photometric uncertainties in our mass estimation we use a Monte Carlo routine, again similar to the one described in \cite{muzic_2019}. In short, we produce a set of 1000 initial magnitudes in the $J$-band, assuming a normal distribution with a standard deviation equal to the respective photometric uncertainty. The extinction correction and conversion to masses is then carried out for each of the 1000 iterations. As a result, we obtain a distribution of 1000 masses for each star.  

\begin{figure*}
    \includegraphics[width=\textwidth]{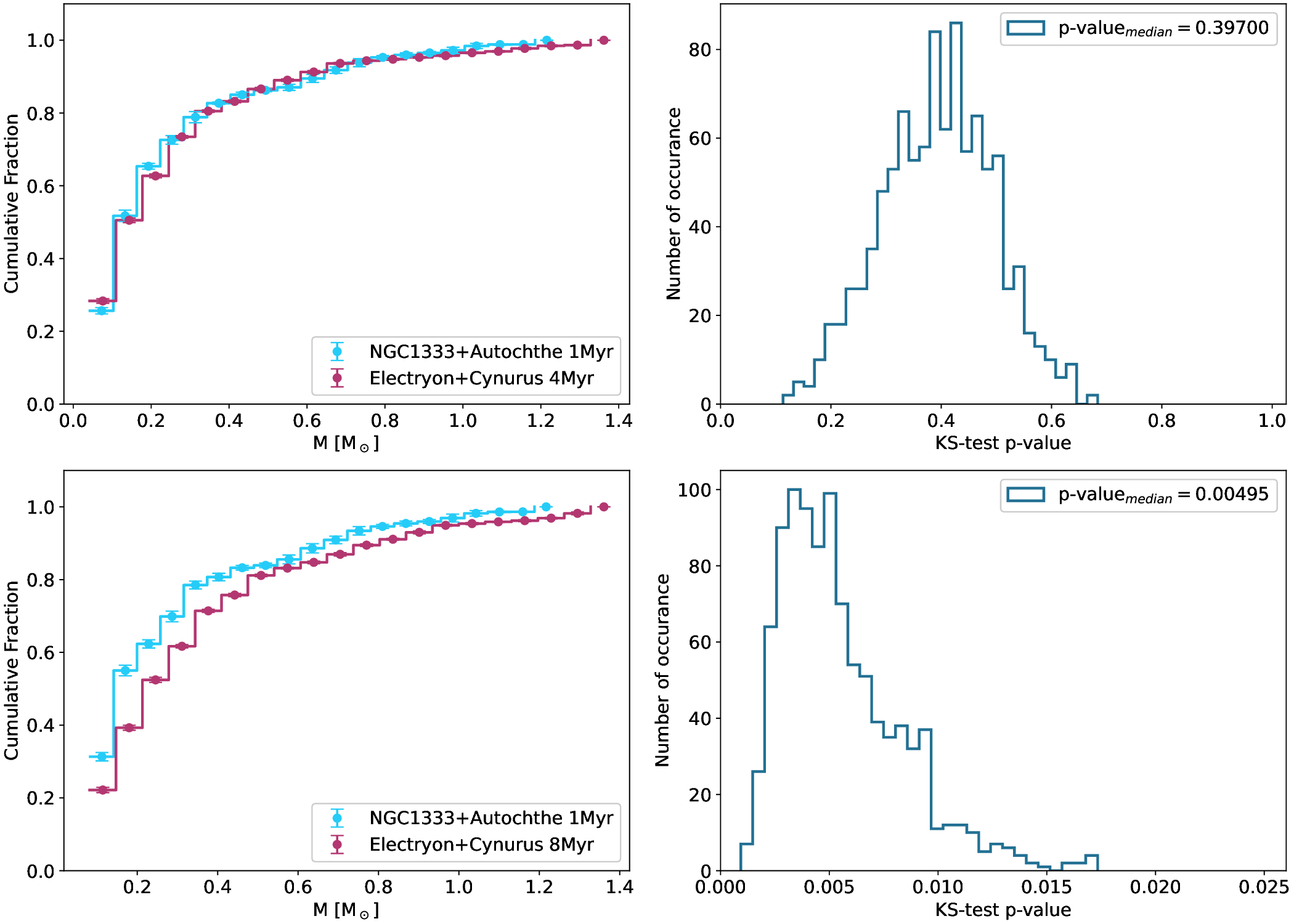}
    \caption{The binned mean cumulative mass distributions of NGC\,1333+Autochthe at 1\,Myr and of Electryon+Cynurus at 4\,Myr (top left) and at 8\,Myr (bottom left). The top and bottom right panels show the histogram of the p-values from the KS-test. Note that the KS tests itself is conducted on unbinned mass distributions. The median p-value which we adopt per comparison is also given for each case in the right panel. The top panels correspond to a case where mass distributions are indistinguishable according to our test. The bottom panels correspond to a case of distinguishable mass distributions. The completeness limits in mass for these cases are $0.03\,M_{\odot}$ (NGC\,1333+Autochthe at 1\,Myr), $0.04\,M_{\odot}$ (Electryon+Cynurus at 4\,Myr), and $0.08\,M_{\odot}$ (Electryon+Cynurus at 8\,Myr).}
    \label{fig:cdf_ngc1_ele4}
\end{figure*}

We conduct another set of Monte Carlo simulations where we randomly draw masses 1000 times from the saved mass distribution per star. We therefore end up with 1000 mass distributions per cluster, per age, which will be analysed further below. We show examples of mean mass distributions derived from this process in Figure \ref{fig:cdf_ngc1_ele4}.

To test the effect of potential binaries on the mass distributions of the clusters, we carry out the whole procedure again, excluding sources that might not be single stars, according to the RUWE criterion discussed above.

The completeness limit in magnitudes (Section \ref{sec:samples}) is converted to the equivalent mass limit adopting an average distance of 300\,pc and an average extinction in $A_V$ of 1\,mags, for each cluster and age. The mass distribution is then truncated at the estimated completeness limit.

\section{Estimating ages from mass distributions}
\label{sec:ages}

\subsection{Comparisons of mass distributions}
\label{sec:KS_test}
Now we turn to the comparisons of the mass distributions. For this process we run Kolmogoroff-Smirnoff (KS) tests using the \textit{ks-2samp} function from \textit{scipy.stats} in \textsc{PYTHON}. Each KS test results in a probability that the two mass distributions are drawn from the same parent distribution. To compare two clusters, we use only the mass range where we have complete information in both clusters, i.e. we only use masses above the completeness limits.

To compare one cluster with another, we run 1000 KS tests, for each pair of mass distributions, which results in a distribution of 1000 probabilities. We adopt the median probability as our final result. If this probability is 0.05 or larger, we accept that the mass distributions are drawn from the same parent distribution. To re-iterate the basic idea of this paper, we assume here that star clusters in Perseus form with a uniform mass distribution. We do not make any assumption about the exact distribution, except that it is uniform across clusters.

In essence, we are searching for the best fitting ages for two clusters, by looking for combinations of ages that produce consistent mass distributions. One caveat of this method is that it may overestimate the absolute p-values -- when varying the ages, the mass distributions change, i.e. we optimize the data to obtain a better fit. This increases the odds of producing a high p-value, by chance. For our purposes, this is not relevant, since we are only interested in the relative p-values, as will become obvious further below.

\begin{figure*}
  \includegraphics[width=\textwidth]{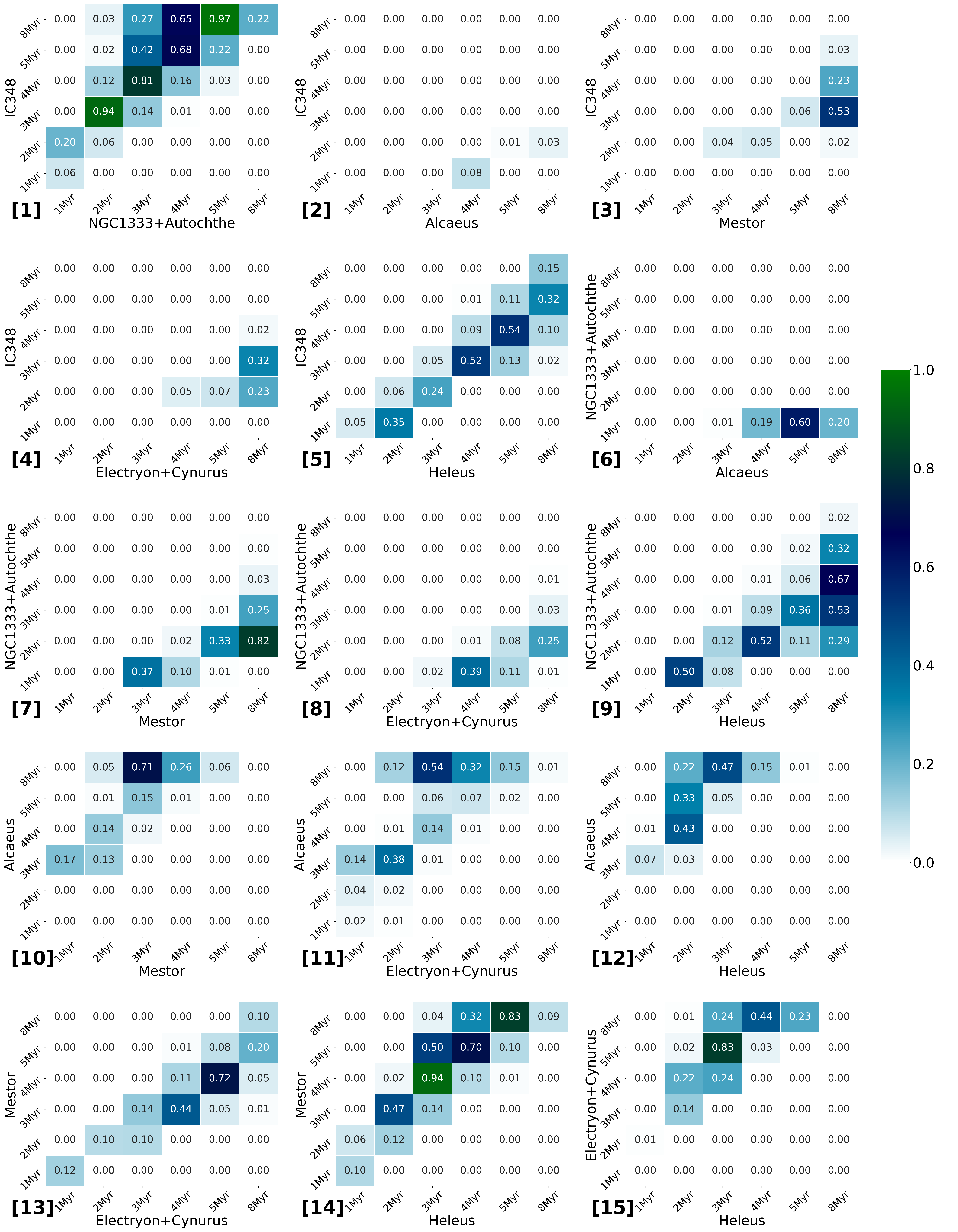}
    \caption{The median probabilities from our KS-tests that the clusters' mass distributions are drawn from the same parent distribution for all cluster and age combinations. The colour-scale also depicts the median probabilities.}
    \label{fig:p_values}
\end{figure*}

\begin{figure*}
\includegraphics[width=0.8\textwidth]{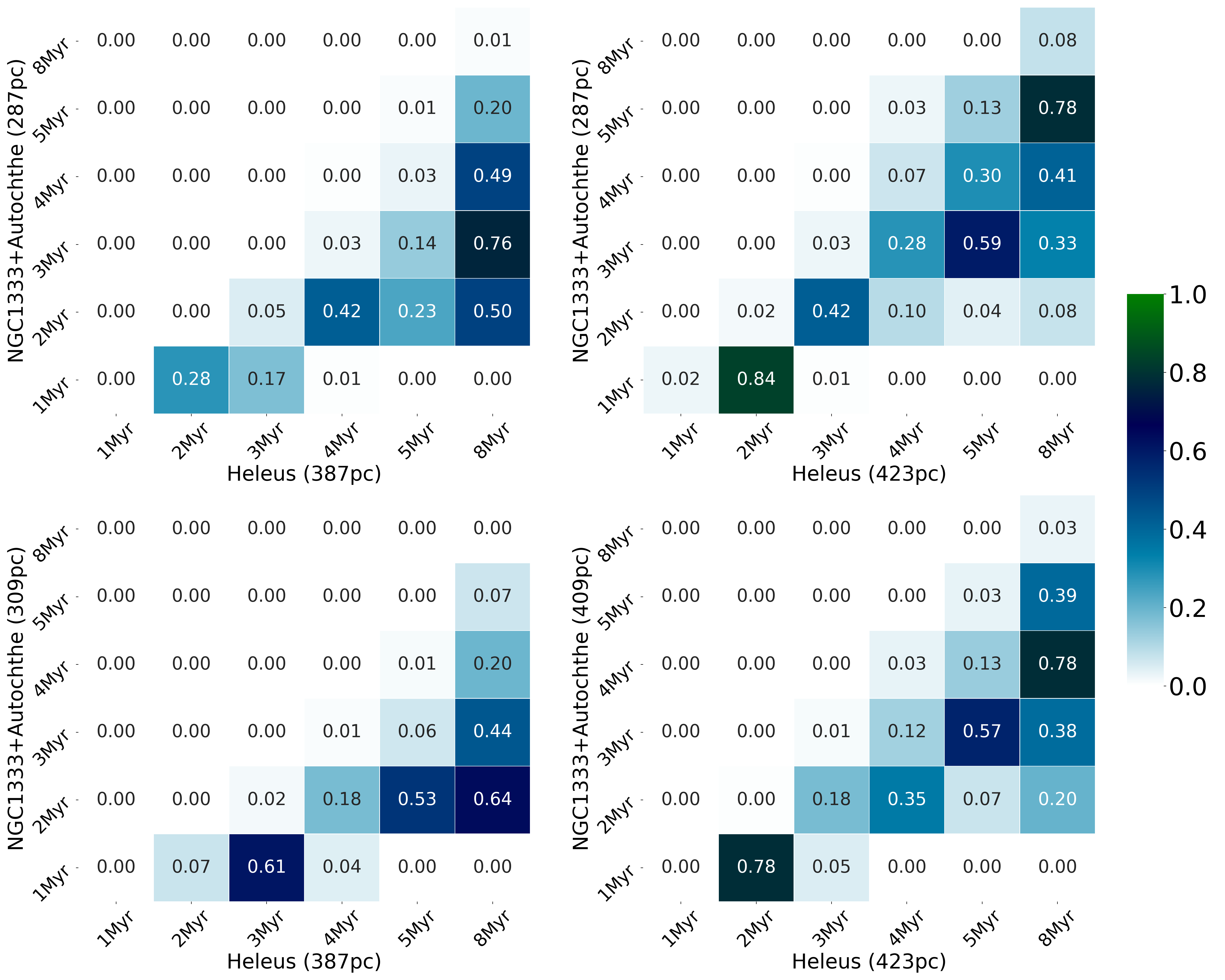}
\caption{The median probabilities that the mass distributions of NGC\,1333+Autochthe and Heleus are drawn from the same parent distribution. As in Figure \ref{fig:p_values} the colour-scale also depicts the median probabilities. The four panels correspond to the four combinations of minimum vs maximum distance within our distance errorbars for the two clusters.} 
\label{fig:p_vals_ngc_hel}
\end{figure*}

As an example product of our methodology, in Figure \ref{fig:cdf_ngc1_ele4} we show the binned mean cumulative mass distribution (left panels) and the histogram of p-values from our KS-test comparison (right panels) for two cases; one where the distributions are similar and one where they are not. Namely, the top panels of Figure \ref{fig:cdf_ngc1_ele4} show the comparison between NGC\,1333 at 1\,Myr and Electryon+Cynurus at 4\,Myr and the bottom panels show the comparison between NGC\,1333+Autochthe at 1\,Myr and Electryon+Cynurus at 8\,Myr. These binned mean cumulative functions are shown for the sake of depicting a comparison of the distributions; for the KS tests we use unbinned mass distributions.

\subsection{Results for all pairs of clusters}
\label{sec:resultsclusters}

With 6 clusters, we need to compare 15 unique pairs of clusters. We conduct this comparison for ages of 1, 2, 3, 4, 5, 8\,Myr for all clusters. This results in a total of 36 comparisons per pair of clusters, or $15\times 36 = 540$ in total. The results from this procedure are visualised in Figure \ref{fig:p_values}. For each pair of clusters we show a $6\times 6$ matrix corresponding to the adopted ages. The average p-values from the KS tests are shown in a colorscale where white corresponds to $p = 0.0$, dark blue to $p=0.5$ and green to $p = 1.0$. A high value here indicates that the mass distributions come from a common parent distribution. As explained before, this is the expected outcome assuming a universal underlying mass function, and therefore all combinations of ages that do not result in a p-value significantly different from zero can be discarded. We note that in some comparisons, the p-values are all low (e.g., panel 2). In others, a wide range of ages yields acceptable p-values (e.g., panel 1). In the following interpretation, we focus mostly on the panels that provide strong constraints on the ages.

Panel 6 shows the comparison between Alcaeus and NGC1333+Autochthe. From this panel, the only ages with acceptable $p>0.05$ are 4-8\,Myr for the former and 1\,Myr for the latter cluster. Adopting those ages, panel 8 shows acceptable p-values for ages of 4-5\,Myr for Electryon+Cynurus. Panel 11 then limits the age range for Alcaeus further to 5-8\,Myr. 

With those constraints in place, comparisons with Mestor (panels 7, 10, 13) only give acceptable p-values for an age of 3-4\,Myr for that cluster. Similarly, Heleus is found to give a good fit for ages of 2-3\,Myr (panels 9, 12, 15). Taken all these results together, the only age for IC348 that gives acceptable p-values is 2\,Myr (panels 1, 3, 4, 5).

In summary, the age sequence for the Perseus star forming region according to our evaluation is NGC\,1333+Autochthe,  IC\,348, Heleus, Mestor, Electryon+Cynurus, Alcaeus. The ages of IC348, Heleus, Mestor, and Electryon+Cynurus are overlapping and cannot reliably be distinguished. We emphasize that, in Figure \ref{fig:p_values} we frequently observe a diagonal pattern in the panels. This means that the age differences between clusters are clearly discernible, even in cases where the absolute ages are difficult to narrow down.  We note that these outcomes do not depend on the absolute p-values; we only select combinations of ages that yield p-values significantly different from zero.

\begin{figure*}
\includegraphics[width=\textwidth]{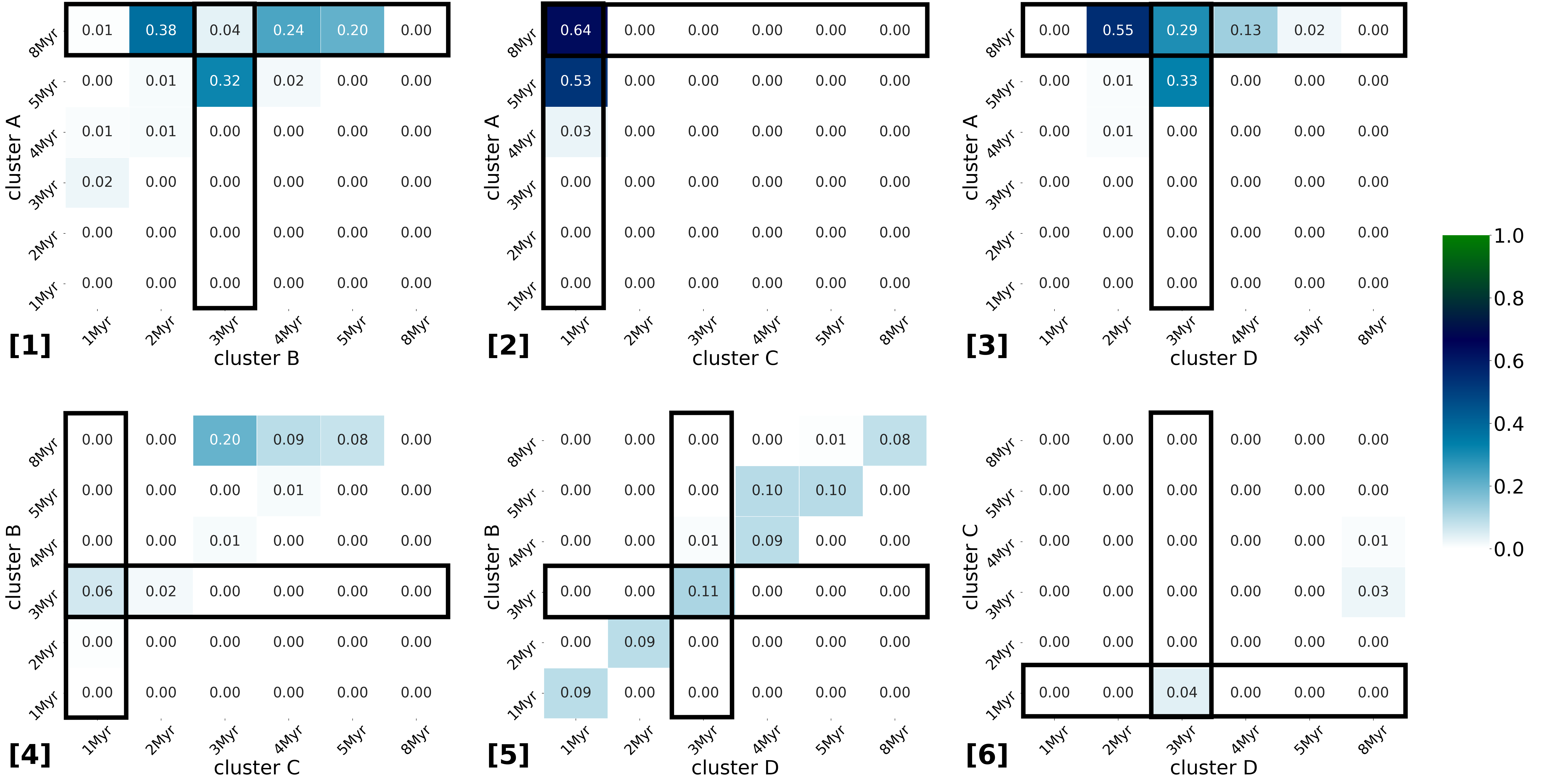}
\caption{The median probabilities that the mass distributions of the four artificial clusters are drawn from the same parent distribution for all combinations of ages. As in Figure \ref{fig:p_values} the colour-scale also depicts the median probabilities. The selected columns and rows show the true ages of the simulated clusters.}
\label{fig:p_values_fake}
\end{figure*}

To test the effect of the error on distance, D, we repeat our analysis for one example, while varying the distances within the errorbars. In Figure \ref{fig:p_vals_ngc_hel} we show the comparison for NGC\,1333+Autochthe and Heleus (panel 9 in Figure \ref{fig:p_values}) for four combinations of distances, which correspond to all possible combinations of maximum/minimum distances given our errorbars (see Table \ref{tab:cluster_conditions}). As can be appreciated in the figure, the uncertainties in distance will affect the absolute values for the probabilities for a given comparison. In particular, for larger distances we would estimate higher luminosities, and hence younger ages overall. This expected trend is reflected in Figure \ref{fig:p_vals_ngc_hel}. For example, when comparing the top panels, the larger distance for Heleus yields the highest probabilities at slightly younger ages. However, the range of plausible ages for a given cluster is unchanged and robust against variations in the distance.

Finally we re-carry our analysis to test the effect of potential binary sources on our results. As aforementioned we remove sources with RUWE $>1.4$ which removes at most 11\% (Alcaeus) of the clusters' members. Omitting binary sources shows no significant changes in the probabilities and therefore our conclusion of the age sequence is unchanged.

\subsection{A test with artificial clusters}

To examine the validity of our methodology, we apply it to a set of 4 artificial clusters A, B, C, D. To create these artificial datasets, we draw samples of masses from a uniform mass function, imitating the properties of our real clusters. The four artificial clusters have 100-300 members, a minimum mass of 0.04 to 0.07$\,M_{\odot}$, and a maximum mass of 1.4$\,M_{\odot}$. We then estimate photometry in J, H, K for those samples, using again the \citet{baraffe_2015} isochrones with ages ranging from 1 to 8\,Myr. We add in extinction, with mean $A_V$ ranging from 1 to 3\,mag and a (Gaussian) scatter in $A_V$ ranging from 0 to 1\,mag. We assume a constant photometric error of 0.03 and a constant distance of 300\,pc. The adopted ages are 8, 4, 1, 3\,Myr for clusters A, B, C, D, respectively.

These artificial sets of photometry were then fed into the methodology developed for the clusters in Perseus. The results are shown in the form of a heatmap in Figure \ref{fig:p_values_fake}, which is similar to Figure \ref{fig:p_values}. Here we mark the adopted ages with boxes. 

When examining this figure, we again focus on panels which give clear constraints on the ages. In panel 2, the only ages that give $p>0.05$ are 1\,Myr for cluster C and 5-8\,Myr for cluster A. Panel 1 gives acceptable p-values for ages of 2-5\,Myr for cluster B. Similarly, we can constrain the age of cluster D to 2-4\,Myr from panel 3. These age ranges are also consistent with the results in panels 4-6, which on their own are  not conclusive.

Thus, the implied age sequence is clusters C, D, B, A, which is exactly reproducing the adopted age sequence in the artificial datasets. The age ranges we estimate are also consistent with the adopted ages. We note that no fine-tuning to our method or to the sets of artificial clusters was done in this exercise. This test therefore provides an independent validation of our methodology. 

\section{Discussion: Comparing with alternative age estimates}
\label{sec:discussion}

In the previous chapters we have estimated ages for the young clusters in the Perseus star forming complex, based on the assumption that their mass distributions are the same. This has led to a clear age sequence for the clusters. Here we will compare our results with age estimates from disc fractions and from the literature. 

\subsection{Ages from disc fractions}
\label{sec:disc_fractions}

Young stars are surrounded by circumstellar discs which disperse on timescales of a few million years. As a result, the disc fraction -- the fraction of stars hosting a disc in a population -- can be used as a proxy for age. 

\citet{pavlidou_2021} estimate disc fractions for all Perseus clusters except IC\,348 and NGC\,1333. They use an empirical criterion based on the $K-W2>0.5$ to identify sources with discs, motivated by the study of \cite{teixeira_2012}. In \cite{pavlidou_2022} the same analysis is consistently conducted for all clusters. They find disc fractions of $66\pm14$\% and $58\pm30$\% for NGC\,1333 and Autochthe respectively, consistent with a very young age. According to the same study, IC\,348 has a slightly lower disc fraction of $41\pm7$\%. In the clusters Mestor, Electryon and Heleus $24\pm3$\%, $18\pm2$\% and $25\pm5$\% of the stars host discs, respectively. Finally, based on the same work, Alcaeus has the lowest disc fraction  of $14\pm3$\%. 

\begin{figure}
\includegraphics[width=\columnwidth]{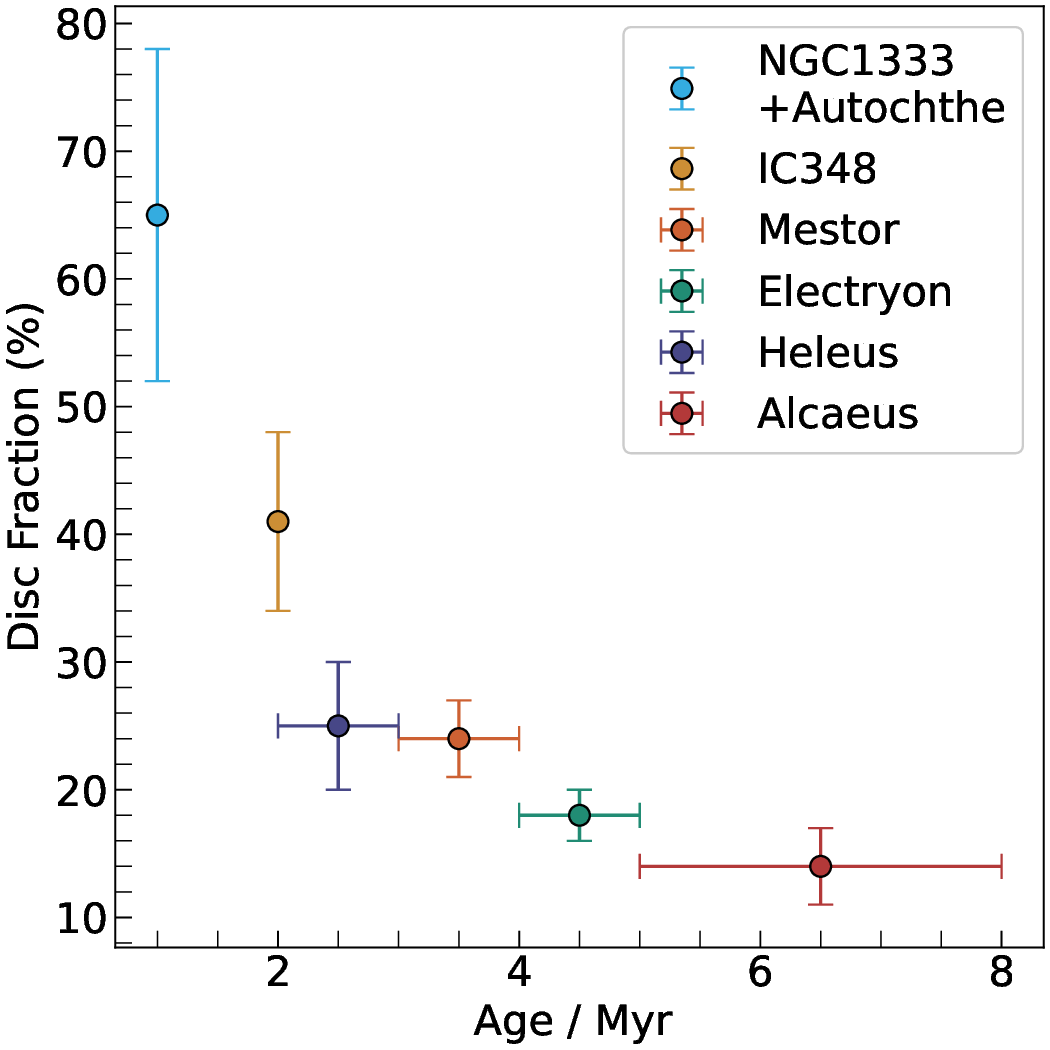}
\caption{Disc fractions of the Perseus clusters as estimated in \citet{pavlidou_2022} from \textit{Gaia} DR2 and \textit{2MASS} data against the age estimates of this work.}
\label{fig:disc_fractions}
\end{figure}

Figure \ref{fig:disc_fractions} shows the disc fractions estimated in \cite{pavlidou_2022} against the ages we establish in this work. Note that in the former, NGC\,1333 and Autochthe are considered separate clusters and Cynurus is not included. In Figure \ref{fig:disc_fractions} we combine NGC\,1333 and Autochthe, shown as one cluster with a disc fraction of $65\pm13$\%. The figure clearly shows the expected trend of declining disk fractions with increasing age. Based on disc fractions the age sequence is NGC\,1333+Autochthe, IC\,348, Heleus, Mestor, Electryon, and Alcaeus. This is entirely consistent with the age sequence derived from the mass distributions in Section \ref{sec:massdist}.  

\subsection{Ages from the literature}

For the two well-studied clusters in this region, the typically quoted ages are 1-3\,Myr for NGC\,1333 and 1-5\,Myr IC\,348 (see the literature review by \citet{olivares_2023}), with a considerable range depending on methodology and sample. \citet{luhman_2016} states that based on the distribution of members in colour-magnitude space, the ages of these two clusters should be similar. For context, we also note that \citet{froebrich2024} find that NGC\,1333 has both a higher disc fraction and a higher fraction of variable members, both indicators of youth. Along similar lines, \citet{gutermuth2009} find a much lower ratio of Class II vs Class I sources  in NGC\,1333 (2.7) compared to IC\,348 (9.0), again a signature of a relatively younger age in NGC\,1333. In any case, the typically assumed ages are certainly consistent with those we estimated from the mass distributions. 

For the more recently identified clusters, the age estimates in the literature are sparse. \cite{kounkel_2022} estimate ages for these groups based on fitting their members samples in colour-magnitude space with theoretical isochrones. Their ages (from their Fig. 3) are 2\,Myr for Autochthe, 2.5\,Myr for NGC\,1333 and IC\,348, 2.8\,Myr for Heleus and Mestor, 3.5\,Myr for Electryon, 4.5\,Myr for Alcaeus, and 7\,Myr for Cynurus. While the overall age sequence is comparable to our findings, there are notable differences. Cynurus, which we have merged with Electryon based on similar kinematics, may be somewhat older than Electryon based on the isochrone fit. However, we note that this estimate relies on very few members on the upper main sequence; the low-mass members of Cynurus would also be consistent with ages of 4-5\,Myr. Their age for NGC\,1333 is slightly older than our estimate. It is notable that \cite{kounkel_2022} introduce a magnitude cutoff at $G=18$ for their Gaia selected sample, which means they will exclude faint, reddenened sources by design and therefore may slightly overestimate the ages for embedded clusters such as NGC\,1333. 

A similar, but independent analysis was conducted by \cite{olivares_2023} for most of the clusters, using various sets of isochrones and samples again selected using Gaia. They find ages of 3\,Myr for NGC\,1333 and Autochthe, 3-5\,Myr for IC\,348 (distinguishing between its core and halo), 5\,Myr for Heleus, 10\,Myr for Alcaeus. Again the sequence of ages is consistent with our results, but the absolute values tend to be somewhat higher. 

For completeness, we add that \citet{pavlidou_2021} also defined an age sequence, with NGC\,1333 and Autochthe being the youngest, followed by Heleus, Mestor, Electryon, and Alcaeus. In this case the ages are inferred from colour-magnitude diagrams. 

Overall, our newly defined age sequence determined by comparing the mass distributions agrees well with literature estimates, validating the methodology proposed in this paper. In terms of absolute ages, the comparison is hampered by the fact that studies typically define their own samples. What is typically measured then, is not the age of the cluster as such, but the age of the population selected using a specific method. When comparing absolute ages, it is therefore necessary to be clear about the sample selection procedure. 

\section{Conclusions}

In this paper we estimate the age sequence for a set of young clusters in the Perseus star forming complex. To this end, we use a new approach: We make use of the fact that the mass-luminosity relation is a strong function of age for pre-main sequence stars, and that the mass function is universal, at least for the star forming regions in the galactic neighborhood. By positing that mass distributions in the studied clusters should be the same, we can estimate ages.

We start by selecting new samples for six young clusters in Perseus, including the two long-known clusters NGC\,1333 and IC\,348, as well as 4 recently identified groups. For the sample selection, we use kinematics from the \textit{Gaia} DR3 catalogue. Following that, we estimate masses by comparing the \textit{2MASS} photometry with theoretical isochrones, while propagating the photometric errors using Monte Carlo simulations. This gives us mass distributions for all clusters, for a set of ages. We confirm our methodology using a set of artificial clusters.

We then compare all combinations of clusters and ages. We select those age combinations that yield (statistically) the same mass distribution. A clear age sequence emerges, with NGC\,1333+Autochthe being the youngest, followed by IC\,348, Heleus, Mestor, Electryon+Cynurus, Alcaeus. This age sequence is consistent with age estimates in the literature using isochrone fitting, as well as with the age sequence inferred from the published disc fractions. It is robust against changes in the distance within the errorbars and is not affected by inclusion or exclusion of suspected binaries. 

In terms of absolute ages, we estimate that NGC\,1333+Autochthe are very young at 1\,Myr, IC\,348 at 2\,Myr, Heleus at 2-3\,Myr, Mestor at 3-4\,Myr, Electryon+Cynurus at 4-5\,Myr, and Alcaeus at 5-8\,Myr. For the most part, these ages are consistent with published values within a margin of $\pm$1-2\,Myr. However, we note that comparing absolute ages is fraught with uncertainty, because they depend critically on the sample selection criteria as well as the choice of the isochrone.

In summary, our paper introduces and validates a new methodology for estimating relative ages of young clusters, which makes use of the established fact that the local mass function is universal.  

\section*{Acknowledgements}

We would like to thank the anonymous referee for a constructive and careful review that helped to improve the paper.
This work has made use of data from the European Space Agency (ESA) mission
{\it Gaia} (\url{https://www.cosmos.esa.int/gaia}), processed by the {\it Gaia}
Data Processing and Analysis Consortium (DPAC,
\url{https://www.cosmos.esa.int/web/gaia/dpac/consortium}). Funding for the DPAC
has been provided by national institutions, in particular the institutions
participating in the {\it Gaia} Multilateral Agreement. 
This research has made use of Python, \url{https://www.python.org}, NumPy \citep{vanderwalt11}, and Matplotlib \citep{hunter07}. This research
made use of TOPCAT (Tool for OPerations on Catalogues And Tables), an interactive graphical viewer and editor for tabular data \citep{taylor05}. TP acknowledges funding support from the Cyprus University of Technology "Postdoctoral Program", 2023 under the acronym "DyMAP". AS acknowledges support from the UKRI Science and Technology Facilities Council through grant ST/Y001419/1/.


\section*{Data Availability}

The data underlying this article are entirely taken from publicly available archives which are described and referenced in the Acknowledgements.




\bibliographystyle{mnras}
\bibliography{ms} 



\appendix
\section{Selection of Cluster Members}
\label{sec:selection_appendix}

In this Appendix we include the selection criteria that we require for the members of each cluster. Table \ref{tab:cluster_conditions} shows the initial spatial selection (box and ellipse), the proper motion selection (box and ellipse), the number of members in our final sample, the cluster's center in the ($\alpha$,$\delta$) plane, and its distance in pc.

\begin{table*}
\centering
\caption{Selection criteria and properties for the Perseus clusters}
\label{tab:cluster_conditions}
\begin{tabular}{lccccccc} 
\hline
\hline
& NGC\,1333 + Autochthe  & IC\,348  & Alcaeus  & Mestor  & Electryon  & Heleus  & Cynurus   \\
\hline
\multicolumn{8}{c}{Box in Spatial Distribution (deg) }  \\
\hline
($\alpha$,$\delta$)$_\mathrm{center}$   & 51.5, 31.0  & 56.0,32.25 & 58.25,32.0  & 57.75,35.25  & 60.5,32.25 & 56.5,29.75  & 61.25,34.25  \\
box half-length                         & 0.8         & 2.0        & 1.6         & 2.0          & 2.5        & 1.5          & 1.0          \\
\hline
\multicolumn{8}{c}{Ellipse in Spatial Distribution (deg)}  \\
\hline
ellipse                                                   & 3$\sigma$              & 3$\sigma$         & 2$\sigma$         & 2$\sigma$         & 2$\sigma$         & 2$\sigma$        & 2$\sigma$       \\
($\alpha$,$\delta$)$_\mathrm{mean}$                        & 51.840, 31.124         & 56.136, 32.155    & 58.357, 32.128    & 57.822, 35.066    & 60.434, 32.247    & 56.5720, 30.148  & 61.295, 34.171  \\
$\sigma(\alpha$,$\delta)$                                  & 0.359                  & 0.538             & 0.785             & 0.844             & 1.185             & 0.540            & 0.484           \\
N$_\mathrm{sources}$                                       & 147                    & 786               & 315               & 371               & 710               & 142              & 92              \\
\hline
\multicolumn{8}{c}{Box in Proper Motion (mas\,yr$^{-1}$) } \\
\hline
($\mu_{\alpha}^*$,$\mu_{\delta}$)$_\mathrm{center}$        & 7.25,-9.75             & 4.25,-6.25        & 6.25,-9.75        & 3.25,-4.25        & 3.75,-5.25        & 2.75,-5.25       & 3.75,-5.25      \\
box half-length                                            & 10.0                   & 10.0              & 3.0               & 5.0               & 4.0               & 5.0              & 5.0             \\
\hline
\multicolumn{8}{c}{Ellipse in Proper Motion (mas\,yr$^{-1}$) } \\
\hline
ellipse                                                    & 3$\sigma$              & 5$\sigma$         & 5$\sigma$         & 6$\sigma$         & 6$\sigma$         & 6$\sigma$        & 5$\sigma$       \\
($\mu_{\alpha}^*$,$\mu_{\delta}$)$_\mathrm{mean}$          & 7.268, -9.644          & 4.451, -6.515     & 6.203, -9.571     & 3.380, -4.412     & 3.628, -5.425     & 2.717, -5.251    & 3.861, -5.446   \\
$\sigma$($\mu_{\alpha}^*$,$\mu_{\delta})$                  & 1.022                  & 0.814             & 0.353             & 0.246             & 0.365             & 0.264            & 0.371           \\
N$_\mathrm{sources}$                                       & 100                    & 549               & 103               & 220               & 323               & 89               & 42              \\ 
\hline
Final sample (N$_\mathrm{sources}$)                        & 100                    & 549               & 103               & 220               & 322$^{**}$          & 89               & 42             \\ 
\hline
($\alpha$,$\delta$)$_\mathrm{center}$ / deg                & 52.04, 31.23           & 56.05, 32.15      & 58.4, 32.07       & 57.88, 35.09      & 60.55, 32.5$^{***}$ & 56.5, 29.93      & --             \\
Distance / pc                                              & 298 $\pm$ 11           & 323 $\pm$ 22      & 294 $\pm$ 17      & 386 $\pm$ 22      & 360 $\pm$ 19$^{***}$& 400 $\pm$ 23     & --             \\
\hline
\hline
\multicolumn{8}{p{\linewidth}}{\footnotesize$*$ To derive the total standard deviation we combine the two values in right ascension and in declination as $\sqrt{\sigma_{\alpha}^2 + \sigma_{\delta}^2}$ }\\
\multicolumn{8}{p{\linewidth}}{\footnotesize$**$ 1 common source found with Cynurus is removed from the Electryon's list of members} \\
\multicolumn{8}{p{\linewidth}}{\footnotesize$***$ Values for Electryon and Cynurus combined as one cluster (364 members in total)} \\
\end{tabular}
\end{table*}

\section{Analysis using the median mass}
\label{sec:p_values_median}

\begin{figure*}
\includegraphics[width=\textwidth]{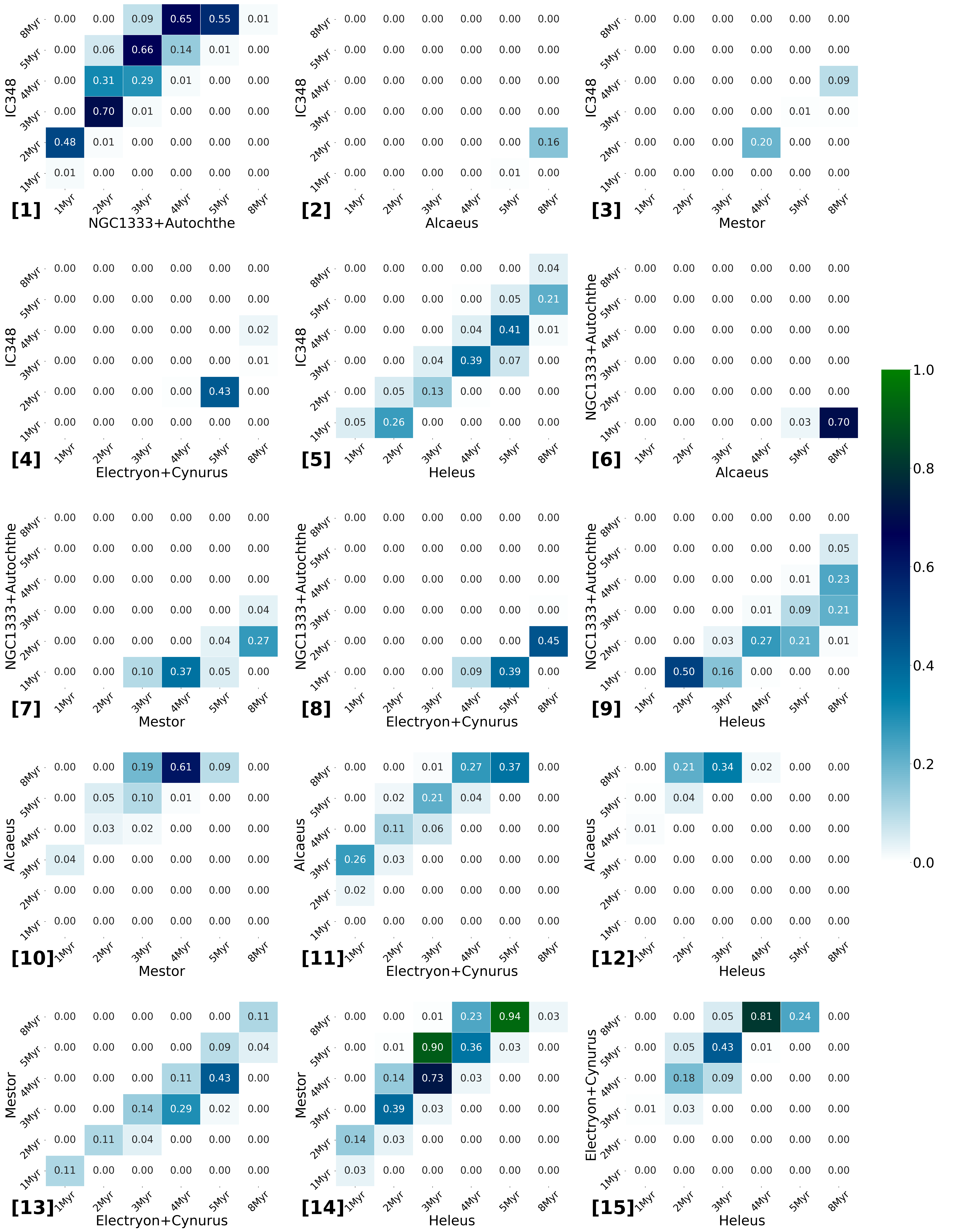}
\caption{The median probabilities that the mass distributions of the Perseus clusters are drawn from the same parent distribution for all combinations using the median mass estimate per cluster member (not the mass distribution per cluster member).}
\label{fig:p_values_mass_med}
\end{figure*}

In this Appendix we reproduce our Figure \ref{fig:p_values} which shows the p-values for all cluster / age comparisons based on the method described in the text. However, instead of using a distribution of masses for each star, we adopt here the median mass, which simplifies the procedure and provides and independent check. This means we are only conducting 1 KS test per cluster / age combination. Figure \ref{fig:p_values_mass_med} shows the corresponding p-values from the KS tests using mass medians.

Examining this figure in the same way as in Section \ref{sec:resultsclusters} yields very similar results. For example, panel 6 limits the age of NGC1333-Autochthe to 1\,Myr. From panel 4 we would infer an age of 2\,Myr for IC348. Panel 12 yields 2-3\,Myr for Heleus. With those constraints, we obtain 4-5\,Myr for Electryon+Cynurus (e.g., panel 8) and 3-5\,Myr for Mestor (e.g., panel 7). For Alcaeus, this test only gives acceptable p-values for an age of 8\,Myr (as opposed to 5-8\,Myr when using mass distributions). In any case, these results are entirely consistent with our established age sequence.

\bsp	
\label{lastpage}
\end{document}